\documentclass[runningheads]{llncs}

\usepackage[T1]{fontenc}
\usepackage{graphicx}
\usepackage{xcolor}

\begin{document}

\title{Accessibility Beyond Accommodations: A~Systematic Redesign of Introduction to Computer~Science~for~Students with~Visual~Impairments}

\titlerunning{Accessible ICS Curriculum}
%
\author{Vaanee Tripathi and Aalok Thakkar}
\authorrunning{}
%
\institute{}
\maketitle              
\begin{abstract}
Computer science education has evolved extensively; however, systemic barriers still prevent students with visual impairments from fully participating. While existing research has developed specialized programming tools and assistive technologies, these solutions remain fragmented and often require complex technical infrastructure, which limits their classroom implementation. Current approaches treat accessibility as individual accommodations rather than integral curriculum design, creating gaps in holistic educational support.
This paper presents a comprehensive framework for redesigning introductory computer science curricula to provide equitable learning experiences for students with visual impairments without requiring specialized technical infrastructure. The framework outlines five key components that together contribute a systematic approach to curriculum accessibility: accessible learning resources with pre-distributed materials and tactile diagrams, in-class learning kits with hands-on demonstrations, structured support systems with dedicated teaching assistance, an online tool repository, and psychosocial support for classroom participation.
Unlike existing tool-focused solutions, this framework addresses both technical and pedagogical dimensions of inclusive education while emphasizing practical implementation in standard university settings. The design is grounded in universal design principles and validated through expert consultation with accessibility specialists and disability services professionals, establishing foundations for future empirical evaluation of learning outcomes and student engagement while serving as a template for broader institutional adoption.

\end{abstract}

\section{Introduction}
\label{sec:intro}

The rapid expansion of computer science education across K-12 and higher education institutions reflects computing's fundamental role in modern society. However, this growth has not translated to equitable participation for all learners. Students with visual impairments face substantial systemic barriers that prevent their full engagement with computer science education and limit their pathways to computing careers \cite{litreview_stefik2019csp}.

\subsection{The Scale of Exclusion}

India alone houses approximately 4.95 million blind and 35 million visually impaired individuals, representing about one-quarter of the world's visually impaired population \cite{intro_mannava2022economic}. This demographic faces severe workforce exclusion, with approximately 9.9 million persons with disabilities in employable age groups classified as non-workers or marginal workers \cite{intro_kumar2017education}.

Educational research reveals systematic discouragement from STEM participation. Studies of students with visual impairments in India document institutional gatekeeping practices where ``it was intensely drilled into the learners with VI that science was not for them'' \cite{intro_palan2020science}. This exclusion begins early and persists throughout educational pathways, creating cumulative disadvantages that extend into professional careers.

\subsection{Infrastructure and Institutional Barriers}

While educational infrastructure for students with visual impairments has expanded: India grew from 4 schools for the blind in 1900 to 973 special education centers by 2016 \cite{intro_kumar2017education}, the quality and accessibility remain problematic. Current data reveals significant gaps: only 65\% of special schools provide inclusive education, 37\% offer vocational training, and merely 16\% provide higher education services. \cite{intro_kumar2017education}

Higher education institutions demonstrate similar accessibility challenges. A comprehensive survey of 185 university libraries found that only 22.07\% provided assistive technology access, with a mere 0.54\% offering tactile pathways for STEM instruction \cite{intro_atprovision}. These infrastructure limitations create compounding barriers that systematically exclude students with visual impairments from computing education \cite{intro_atprovision}\cite{intro_kumar2017education} \cite{intro_palan2020science}.

At the workforce level, the unemployment rate of the legally blind or visually impaired population in the United States hovers near 63\%, with fewer than 6.5\% of employed individuals with any disability working in STEM disciplines \cite{intro_chemistry_camp}. This employment disparity reflects deeper educational inequities that begin early in the academic pipeline. In the eleven states that reported comprehensive data, while 12.9\% of K-12 students were served under the Individuals with Disabilities Education Act (IDEA), only 7.6\% of students who took at least one computer science course were served under IDEA, revealing significant underrepresentation in computer science education \cite{litreview_accessiblecs_whitepaper}.

The exclusion becomes more pronounced at the collegiate level, where 32\% of undergraduate computer science majors with disabilities feel like outsiders in the computing field. This percentage increases dramatically to 45-46\% for women or underrepresented minorities with disabilities, compared to only 17\% for majority men with disabilities \cite{litreview_accessiblecs_whitepaper}. These statistics represent not merely numbers, but lost potential among the approximately 1.3 million Americans who are considered legally blind and the 1.5 million visually impaired computer users \cite{intro_ludi2011robotics}.

\subsection{Curricular Accessibility Challenges}

Standard computer science curricula present fundamental accessibility barriers that extend beyond infrastructure limitations. Visual programming environments like Alice, Scratch, Blockly, and Turtle that are designed specifically to attract young learners to computing remain largely inaccessible to blind students \cite{scitepress2020}. These platforms rely heavily on visual metaphors and drag-and-drop interfaces that are incompatible with screen readers, effectively excluding students from mainstream introductory programming experiences \cite{intro_ludi2011robotics,intro_milne2018blocks4all}.

The reliance on visual programming creates a pedagogical paradox: tools designed to increase accessibility and engagement for sighted novice programmers simultaneously create barriers for students with visual impairments. This forces students to choose courses based on accessibility constraints rather than ability and interest, limiting educational and career opportunities \cite{intro_2020systematic} \cite{intro_palan2020science}. 

\subsection{Limitations of Current Approaches}

Existing approaches to accessibility in computer science education remain fragmented and insufficient \cite{intro_mountapmbeme2022addressing}. While assistive technologies exist, implementation requires complex technical infrastructure that limits practical classroom deployment \cite{intro_mountapmbeme2022addressing}. Current research tends to focus on individual tools or specific accommodations rather than comprehensive curricular solutions (outlined further in the review of literature), treating accessibility as an add-on rather than an integral design consideration.
This accommodation-based approach creates several problems: it places the burden of accessibility on individual students and instructors, results in delayed or incomplete access to course materials, and often provides alternative rather than equivalent learning experiences. These limitations highlight the need for systematic approaches that embed accessibility into curricular design from the outset.

\subsection{Research Question and Contributions}

The persistent exclusion of students with visual impairments from computing education, despite decades of assistive technology development, suggests fundamental limitations in current pedagogical approaches. This work investigates the following research question: \textit{How can introductory computer science curricula be systematically redesigned to ensure equitable learning outcomes for students with visual impairments while maintaining pedagogical effectiveness and practical implementation feasibility?}

We present an empirically grounded framework that integrates accessibility considerations into core curricular design rather than treating them as peripheral accommodations. Our approach synthesizes universal design for learning principles with evidence-based practices in computing pedagogy to create five interconnected components: (1) multi-modal learning resources designed for screen reader compatibility, (2) tactile programming kits that support hands-on algorithmic thinking, (3) peer mentoring and instructor support structures, (4) accompanying resources for tools and course-content, and (5) scaffolded psychosocial interventions addressing stereotype threat and self-efficacy.

The framework extends beyond immediate accessibility to prepare students for career success by emphasizing collaborative learning and authentic assessment methods that mirror professional software development practices. Students gain experience with assistive technologies and inclusive team dynamics that directly translate to workplace environments, potentially addressing the documented workforce exclusion experienced by individuals with visual impairments and improving employment outcomes in computing fields.

The framework can be validated through panel reviews, one-on-one interview sessions with students with visual impairments, and a pilot semester for the curriculum. Our work demonstrates that inclusive pedagogical design can improve learning outcomes for all students while eliminating systemic barriers for those with visual impairments. This work shifts the accessibility paradigm from retrofitted accommodations toward proactive universal design, providing a scalable model for inclusive computer science curriculum development.
\section{Review of Literature}
\label{sec:review}

\subsection{Accessibility Barriers in Mainstream Computing Education}

Contemporary CS education relies heavily on visual programming paradigms that systematically exclude students with visual impairments. Stefik et al. conducted a comprehensive accessibility audit revealing that {\em all endorsed AP Computer Science Principles curricula are not fully accessible,} with widely-adopted platforms like Code.org's App Lab demonstrating complete incompatibility with screen reader technology despite deployment across thousands of educational institutions \cite{litreview_stefik2019csp}. This finding highlights a fundamental tension between pedagogical innovation and inclusive design in computing education.

Block-based programming environments, designed to lower entry barriers for novice programmers, paradoxically create insurmountable obstacles for students with visual impairments. Milne and Ladner's systematic analysis demonstrates that these environments {\em rely heavily on visual metaphors and are therefore not fully accessible for children with visual impairment} \cite{intro_milne2018blocks4all}. Their work identifies five critical accessibility barriers: (1) inability to access block semantic information without visual inspection, (2) cognitive overhead in navigating complex block hierarchies, (3) incompatibility of drag-and-drop interaction paradigms with assistive technologies, (4) lack of non-visual mechanisms for distinguishing block categories and types, and (5) absence of accessible feedback systems for block connection validation \cite{intro_milne2018blocks4all}. These barriers emerge from fundamental design assumptions that prioritize visual affordances over multi-modal accessibility.

\subsection{Existing Accessibility Solutions and Their Limitations}

Current research has produced three primary categories of accessibility interventions, each addressing specific aspects of the broader challenge while introducing implementation constraints that limit classroom adoption:

\begin{enumerate}
    \item \textbf{Specialized Programming Environments:} Audio-based programming interfaces represent one approach to accessibility. Francioni and Smith's JavaSpeak system provides eight hierarchical levels of auditory code rendering, enabling non-visual program comprehension through structured speech synthesis \cite{litreview_francioni2002javaspeak}. However, these specialized environments operate in isolation from mainstream educational practice, requiring students to learn alternative paradigms that may not transfer to industry-standard development.
    
    \item \textbf{Tangible and Haptic Interfaces:} Physical programming representations offer another accessibility pathway. Kane and Bigham's STEM×Comet workshop demonstrated Ruby programming instruction using 3D-printed tactile data visualizations, showing promise for hands-on algorithmic thinking \cite{litreview_kane2014stemxcomet}. However, tangible solutions encounter significant practical constraints, including lighting dependencies for mixed-reality systems, complex hardware setup requirements, as well as initial capital investment and maintenance overhead unsuitable for universal classroom deployment \cite{intro_mountapmbeme2022addressing}.
    
    \item \textbf{Robotics-Based Pedagogies:} Educational robotics provides a concrete, multi-sensory programming experience. Ludi and Reichlmayr's JBrick system offers accessible text-based interfaces for Lego Mindstorms programming, enabling students with visual impairments to participate in robotics curricula \cite{intro_ludi2011robotics}. While robotics-based approaches show pedagogical promise, they require substantial equipment investment, ongoing technical maintenance, and specialized instructor training that limit scalability.
\end{enumerate}

\subsection{Systemic Gaps in Accessibility Research}

A critical analysis of existing accessibility research reveals fundamental limitations in scope and approach. Most interventions focus on individual tool development rather than systematic curriculum transformation. This tool-centric perspective treats accessibility as an accommodation to be retrofitted rather than an integral design consideration from the outset. Even comprehensive efforts like Stefik et al.'s AP CSP curriculum modifications only address specific programming environments without extending to broader curricular integration, where accessibility challenges compound with increased technical complexity \cite{litreview_stefik2019csp}.

Furthermore, existing solutions frequently require additional time investment beyond standard course requirements for students with visual impairments to achieve equivalent learning outcomes \cite{intro_mountapmbeme2022addressing}. This creates inequitable learning experiences that may discourage persistence in computing fields. The accommodation paradigm also places a disproportionate burden on individual instructors to implement accessibility measures without institutional support or training.

\subsection{Gaps in Pedagogical Integration}

Several critical gaps persist in accessibility research for computing education. First, limited attention has been paid to instructor preparation and professional development for inclusive teaching practices. Second, research on collaborative learning experiences for students with visual impairments remains underdeveloped, despite collaboration being central to contemporary computing pedagogy. Third, practical implementation strategies for integrating accessibility into standard classroom environments without specialized technical infrastructure have received insufficient attention \cite{intro_mountapmbeme2022addressing}.

While significant progress has been made in developing accessible programming tools and individual interventions, fundamental gaps remain in creating systematically inclusive computing education that embeds accessibility throughout the pedagogical experience rather than treating it as a peripheral concern.
\section{Framework Design Specifications}
\label{sec:design}

The development of an inclusive introductory computer science curriculum requires systematic consideration of both technical constraints and pedagogical objectives. This section establishes the design specifications that guided framework development, grounded in empirical evidence from accessibility research and established principles of computing education. 

\subsection{Functional Requirements}

Framework development was constrained by four categories of functional requirements derived from analysis of existing accessibility barriers and institutional implementation contexts:

\begin{enumerate}
    \item \textbf{Accessibility Requirements:} The framework must ensure complete compatibility with contemporary screen reading technologies (JAWS, NVDA, VoiceOver) across all learning materials and interactive components. All visual content requires semantically equivalent tactile or auditory alternatives that preserve pedagogical intent. The system must accommodate diverse assistive technology configurations without requiring specialized hardware beyond standard university computing resources.
    \item \textbf{Pedagogical Requirements:} Curricular components must align with established computer science learning objectives as defined by ACM/IEEE Computer Science Curricula guidelines \cite{acm_ieee_cs2023}. Academic rigor must remain equivalent to traditional introductory computer science courses, ensuring that accessibility accommodations do not compromise learning outcomes or career preparation. The framework must integrate with established pedagogical methodologies, including active learning, collaborative programming, and scaffolded skill development.
    \item \textbf{Implementation Requirements:} Solutions must demonstrate feasibility within standard university environments without requiring specialized technical infrastructure or significant investment. Instructor preparation time must remain comparable to traditional course development, with clear documentation and training materials provided. The cost structure must support institutional adoption at scale without external funding dependencies.
    \item \textbf{Social Integration Requirements:} The framework must facilitate meaningful participation in collaborative learning activities, enabling equitable peer interactions and group project engagement. Classroom dynamics should promote universally inclusive environments that benefit all students while avoiding stigmatization of accessibility accommodations.
\end{enumerate}

\subsection{Design Principles}
The framework architecture is grounded in four evidence-based design principles that address both accessibility challenges and broader pedagogical effectiveness:

\begin{enumerate}
    \item \textbf{Universal Design for Learning Integration:} Following Rose and Meyer's UDL framework, the curriculum provides multiple means of engagement through varied motivational approaches, multiple means of representation through diverse content modalities, and multiple means of action and expression through flexible assessment options \cite{design_understood_udl}. This approach benefits all students rather than serving as specialized accommodations, reducing stigmatization while creating universally effective learning experiences \cite{design_udhe_doit}.
    \item \textbf{Progressive Scaffolding:} Learning experiences follow Vygotsky's zone of proximal development theory, with carefully structured progression from foundational concepts to complex programming paradigms. This scaffolding approach ensures that students develop robust conceptual understanding before encountering advanced accessibility tool interactions, reducing cognitive load and supporting transfer to professional development environments \cite{vygotsky}.
    \item \textbf{Multi-modal Content Architecture:} Information presentation combines auditory, tactile, and textual modalities following principles of cognitive load theory. This approach ensures {\em perceptible information} accessibility while providing {\em flexibility in use} through multiple access pathways that accommodate diverse learning preferences \cite{design_understood_udl}.
    \item \textbf{Embedded Accessibility Design:} Rather than retrofitting accessibility features onto existing curricula, accessibility considerations are integrated throughout the design process. This approach ensures {\em equitable use} and maintains {\em simple and intuitive} learning experiences for all students while avoiding the accommodation paradigm that places additional burden on individual instructors.
\end{enumerate}

\subsection{Technical Constraints}

Several technical constraints shaped framework development based on analysis of institutional computing environments and accessibility technology limitations:

\begin{enumerate}
    \item \textbf{Screen Reader Compatibility:} All interactive components must function effectively with major screen reading platforms without requiring custom configuration or specialized training. This constraint necessitated careful attention to semantic markup, keyboard navigation patterns, and auditory feedback design.
    \item \textbf{Infrastructure Independence:} Solutions cannot depend on specialized hardware, proprietary software, or high-bandwidth network connections that may not be available in standard classroom settings. This constraint favored web-based tools and open-source technologies over custom applications requiring installation or maintenance.
    \item \textbf{Scalability Requirements:} Framework components must support simultaneous use by multiple students without performance degradation or resource conflicts. This constraint influenced architectural decisions regarding server-side processing, client-side caching, and collaborative tool design.
\end{enumerate}

These specifications collectively ensure that the resulting framework addresses both the immediate accessibility needs of students with visual impairments and the broader institutional requirements for sustainable, scalable computing education reform.
\section{Curriculum Redesign Framework}
\label{sec:curriculum}

We have developed a curriculum redesign framework that addresses systemic barriers identified in traditional computer science education by restructuring the learning experience into five integrated components targeting specific aspects of inclusive classroom learning while maintaining rigour and practical implementation capability.

\subsection{Lecture Materials}
Traditional lecture delivery creates significant barriers for students with visual impairments, particularly when instructors rely on spontaneous visual demonstrations or unplanned code examples. Our framework addresses these challenges through systematic material preparation and alternative presentation strategies.

We provide all lecture content, including code examples, diagrams, and visual demonstrations, to all students before class sessions. This advance distribution enables students to familiarize themselves with complex material using preferred assistive technologies and eliminate exclusion occurring when instructors present impromptu content. Additionally, we have incorporated a {\em Book of Diagrams}: a collection of tactile diagrams corresponding to standard visual content that the course will cover. This resource functions as an appendix that students can reference routinely, allowing instructors to direct students to specific diagrams when teaching visual concepts. These diagrams could be produced using tactile printers or affordable and locally available materials like thread and paper.

Software selection for teaching material preparation requires careful evaluation to ensure accessibility compatibility. We prepare all lecture materials using LaTeX to ensure mathematical expressions, code formatting, and technical notation remain accessible when converted to alternative formats. Our visual presentations undergo a comprehensive accessibility conversion, including detailed alt-text descriptions for all images, diagrams, and code snippets. Material preparation follows best practices, including adding metadata, using heading hierarchy, alt-text, high contrast colours, descriptive links, and lists where possible, while avoiding reliance solely on visual indicators \cite{design_harvard_powerpoint_accessibility}.

This systematic approach transforms ad-hoc visual demonstrations into structured, accessible learning experiences, benefiting all students while ensuring equitable access to course content.

\subsection{Online Tools}
To enable meaningful participation, we recognise that students should achieve proficiency with tools, reducing dependency on visual interfaces, and be provided with comprehensive resources to access and utilize them effectively. Our resource provision is organized into technical aids and conceptual resources for multimodal access.

Students develop proficiency with platform-specific screen-reading technologies through our framework. For Windows users, we identified NVDA as providing free, open-source screen-reading capabilities with coding optimizations. MacOS users benefit from built-in VoiceOver functionality. We selected Visual Studio Code as the recommended IDE due to robust screen reader support and extensive accessibility features. Our framework integrates Maidr (Multi-modal Access and Interactive Data Representation), providing non-visual access to statistical plots through braille, text, and sonification modes. Programming language mastery requires accessible syntax references and documentation for OCaml and Python, which we compiled from official documentation.

Rather than presenting these as supplementary accommodations, we integrate them into standard course materials, ensuring all students benefit from multi-modal learning approaches while reducing stigma.

\subsection{Learning Kit}

We built a learning kit that transforms abstract computer science concepts into tangible, multi-sensory experiences, enabling all students to engage with foundational programming concepts through tactile and auditory modalities.

The kit includes specialized Braille playing cards for algorithmic understanding and 3D-printed apparatus to teach linked lists, control flow, recursion, and iteration concepts. Playing cards feature both Braille and large print markings, facilitating collaboration between visually impaired and sighted students. Each card represents data elements physically manipulated to demonstrate sorting algorithms, list operations, and search procedures. Cards are paired with tactile markers and textured clips for position marking.

\begin{figure}[t]
\begin{center}
\includegraphics[width=1\textwidth]{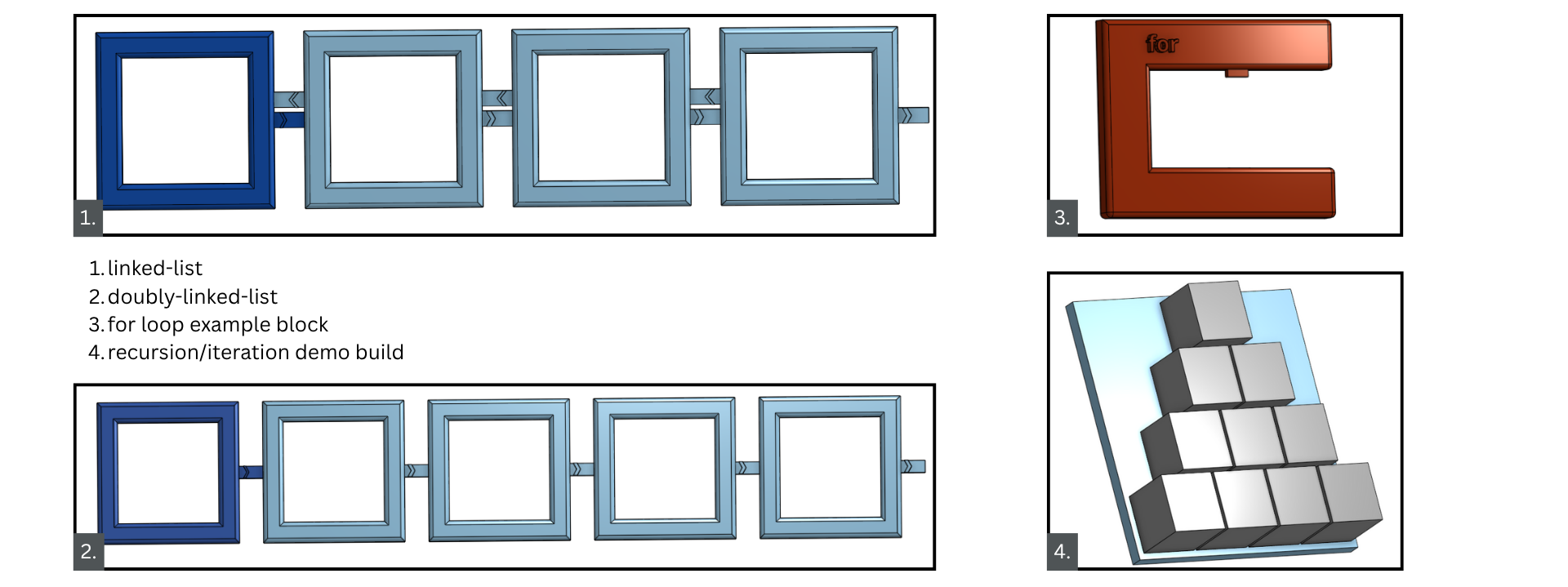}
\end{center}
    \caption{Physical apparatus designs for teaching computer science concepts through tactile learning. The figure shows four different 3D-printed components: (1) Linked-list representation featuring a dark blue head node connected to three light blue nodes via directional arrows, demonstrating unidirectional data flow; (2) Doubly-linked-list representation with a dark blue head node connected to four light blue nodes via bidirectional arrows, showing two-way traversal capability; (3) For-loop control structure block in orange/red color with "for" text label and C-shaped design representing code block enclosure; (4) Recursion/iteration demonstration apparatus consisting of stacked geometric blocks in varying shades (white, gray, and black) arranged in a pyramid formation on a light blue base, designed to physically demonstrate the building and unwinding of recursive function calls versus iterative accumulation patterns.}
    \label{fig:enter-label}
\end{figure}

Demonstrations begin with simplified three-to-five element examples before scaling to realistic problem sizes, ensuring concept mastery before introducing computational complexity. Activities leverage diverse student strengths, creating roles where visually impaired and sighted students contribute equally to problem-solving processes.

\subsection{Teaching Support}
Effective, inclusive education requires systematic support structures extending beyond individual instructor efforts. This component establishes comprehensive training programs and dedicated personnel, ensuring consistent, high-quality, accessible instruction.

We ensure a dedicated teaching assistant specifically trained in accessibility support for students with visual impairments, developed in partnership with the Office of Learning Support at our institution. The dedicated TA selection process prioritizes interpersonal competencies alongside technical knowledge, including demonstrated comfort with disability interactions and genuine enthusiasm for inclusive education.

All teaching personnel participate in mandatory accessibility training covering fundamental principles of inclusive education, addressing common misconceptions about visual impairment, and establishing protocols for respectful interaction. Beyond universal training, dedicated accessibility TAs receive intensive preparation covering screen reader operation, tactile material preparation, and specific techniques for supporting programming instruction.

\subsection{Psychosocial Support}
The psychosocial dimension of inclusive computer science education addresses social dynamics, peer interactions, and emotional well-being factors that significantly impact learning outcomes. This component recognizes that academic accessibility extends beyond technical accommodations to encompass the full social experience of university education. Our approach employs strategic classroom positioning that facilitates natural peer interaction while ensuring optimal access to instruction, avoiding the stigmatization that can occur when accommodation needs visibly separate students from their peers. Group assignments are structured to leverage diverse skills and perspectives, with roles highlighting competencies such as code review, logical problem-solving, documentation, and testing. Extra-credit opportunities feature multiple completion modalities, ensuring students with visual impairments can participate in enrichment activities without encountering accessibility barriers. The framework emphasizes that academic accomplishments by students with visual impairments should receive appropriate recognition without being framed as ``inspirational'' narratives, maintaining dignity while acknowledging academic excellence within the same contexts and criteria applied to all students.
\section{Discussion}
\label{sec:discussion}

Our framework represents a comprehensive design proposal building upon established universal design principles and accessibility research. While grounded in theoretical foundations and expert-informed design principles, the framework requires systematic empirical validation through implementation and testing in real classroom environments. We are implementing a preliminary version and developing prototype learning materials and assessment instruments.

Validation encompasses multiple approaches to assess theoretical soundness and practical feasibility. Ongoing efforts include multidisciplinary expert panel reviews with accessibility specialists, computer science educators, disability services professionals, and students with visual impairments. We are conducting individual consultation sessions with students who have visual impairments for hands-on evaluation of learning kit prototypes and preparing for pilot semester implementation to collect systematic effectiveness data.

The framework acknowledges that students with visual impairments who navigate multiple marginalized identities—including gender, race, ethnicity, or socioeconomic status—may face compounded barriers extending beyond visual accessibility challenges. Successful implementation requires ongoing attention to how intersecting identities influence classroom dynamics and educational experiences.

\subsection{Implementation Challenges and  Limitations}
The framework requires substantial initial investment in tactile materials, assistive technology infrastructure, and comprehensive training programs, presenting significant limitations affecting its generalizability. The design assumes access to institutional resources, including 3D printing capabilities, dedicated Office of Learning Support services, specialized teaching assistant positions, and flexible budgets. These assumptions reflect a high-resource university context and may not translate to institutions with limited accessibility infrastructure or constrained budgets. Training needs extend beyond individual courses to encompass department-wide cultural transformation, while the rapid evolution of assistive technologies necessitates continuous updates to maintain compatibility. Most critically, the framework lacks empirical validation through systematic implementation and assessment of learning outcomes, student satisfaction, or long-term educational impact.

\subsection{STEM Application Potential}

Despite implementation challenges, our preliminary development demonstrates the framework's potential for broader application. Our approach to multi-modal learning and universal design principles is adaptable across STEM disciplines. Mathematics courses could benefit from similar tactile demonstration techniques, while physics and chemistry education could incorporate comparable hands-on learning kits, making abstract concepts accessible through multiple sensory modalities.

Early prototyping reveals both opportunities and constraints in cross-disciplinary adaptation. Each discipline presents unique conceptual challenges and established pedagogical traditions requiring careful consideration during framework adaptation. This comprehensive approach represents a significant departure from traditional accommodation models, moving toward proactive, inclusive design that benefits all learners while addressing the needs of students with visual impairments, offering systematic approaches to inclusive STEM education that could reshape accessibility conceptualization in higher education.
\section{Conclusion and Future Work}
\label{sec:conclusion}

This paper presents a comprehensive framework for redesigning introductory computer science curricula to provide equitable learning experiences for students with visual impairments without requiring specialized technical infrastructure. The five-component approach addresses both technical and pedagogical dimensions of inclusive education while emphasizing practical implementation in standard university settings, fundamentally transforming computer science education by replacing reactive accommodation with proactive inclusive design.

Our preliminary development demonstrates the framework's effectiveness and scalability. Through consultation with students with visual impairments and accessibility experts, we are establishing evidence-based practices that enhance learning outcomes for all students while eliminating historical barriers to computing careers. Our research agenda includes comprehensive longitudinal assessment and cross-disciplinary adaptations.

This work establishes a new paradigm for accessible STEM education that moves beyond compliance toward excellence in inclusive design, providing immediate, actionable solutions for the persistent under-representation of students with visual impairments in computing fields while creating institutional capacity for sustained inclusive education.

\bibliography{references}
\end{document}